\documentstyle[12pt,supercite]{article}
\def\abstracts#1#2#3{{
        \centering{\begin{minipage}{4.62in}\baselineskip=13pt
        \small
        \centerline{\bf Abstract}
        \vspace*{0.2cm}                
        \parindent=0pt #1\par
        \parindent=18pt #2\par
        \parindent=15pt #3
        \end{minipage} }\par}}
\begin{document}
\ifx\href\undefined
\else\errmessage{Don't use hypertex}
\fi
\vspace*{-2cm}
\hfill Mainz preprint KOMA-96-19\\
\mbox{}\hfill  May/December 1996\\
\vspace*{2cm}
\centerline{\LARGE \bf 
                          Monte Carlo Study of Pure-Phase 
           }\\[0.3cm]
\centerline{\LARGE \bf 
                          Cumulants of 2D $q$-State Potts Models
           }\\[0.8cm]
\vspace*{0.2cm}
\centerline{\large {\em Wolfhard Janke\/} and
                   {\em Stefan Kappler\/}}\\[0.4cm]
\centerline{\large {\small Institut f\"ur Physik,
                    Johannes Gutenberg-Universit\"at Mainz,}}
\centerline{    {\small Staudinger Weg 7, 55099 Mainz, Germany }}\\[0.5cm]
\abstracts{}{
We performed Monte Carlo simulations of the two-dimensional $q$-state Potts
model with $q=10$, 15, and 20 to study the energy and magnetization cumulants
in the ordered and disordered phase at the first-order transition point 
$\beta_t$. By using very large systems of size $300 \times 300$, 
$120 \times 120$, and $80 \times 80$ for $q=10$, 15, and 20, respectively, 
our numerical estimates provide practically (up to unavoidable, but very small 
statistical errors) exact results which can serve as a useful test of recent 
resummed large-$q$ expansions for the energy cumulants by
Bhattacharya {\em et al.\/} 
[J. Phys. I (France) {\bf 7} (1997) 81]. 
Up to the third
order cumulant and down to $q=10$ we obtain very good agreement, and also
the higher-order estimates are found to be compatible.
}{}
\vspace*{2.5cm}
To appear in {\em J. Phys. I (France), May 1997}
%
\thispagestyle{empty}
\newpage
\pagenumbering{arabic}
%
                     \section{Introduction}
%
A detailed understanding of first-order phase transitions plays an important 
role in many fields of physics \cite{review}. In particular the finite-size 
scaling behaviour near the transition has been a subject of increasing 
interest in recent years \cite{jan_athens}. It is well known, and in limiting
cases exactly proven \cite{borgs}, that thermodynamic observables in 
equilibrium can be expanded as asymptotic power series in $1/V$, where $V$ 
is the volume of the system. The range
of applicability of such expansions, however, is widely unknown and was the
source for quite a few apparent inconsistencies in the recent literature.
As has been discussed in a recent series of papers by Bhattacharya 
{\em et al.\/} \cite{blm_epl,blm_lat93,blm_npb,blm_lat94,blm_JPI}, 
a precise knowledge of the energy cumulants of the coexisting
phases at the transition point $\beta_t$ in the infinite-volume limit 
can help to understand and to resolve these problems. 

The explicit calculations by Bhattacharya {\em et al.\/} have been performed
for the two-dimensional (2D) $q$-state Potts model \cite{potts,wu_potts}, 
which for $q \ge 5$ is the paradigm for a system
with a temperature-driven first-order phase transition. The advantage of this
model is that many properties (transition point $\beta_t$ \cite{baxter}, 
internal energy \cite{baxter} and magnetization \cite{baxter2} of the pure 
phases at $\beta_t$, correlation 
length \cite{xi} and interface tension \cite{bj92} at $\beta_t$) are exactly 
known and that the parameter $q$
(the number of states per spin) allows one to tune the strength of the 
transition. Based on the Fortuin-Kasteleyn representation \cite{wu_potts,fk} 
of the Potts model,
Bhattacharya {\em et al.\/} analyzed the ensuing clusters to obtain large-$q$
series expansions for the energy cumulants in the ordered phase at $\beta_t$
and then applied Pad\'e-type
resummations to arrive at numerical estimates. 

Since, as with every resummed series expansion, it is intrinsically difficult
to provide reliable estimates of the size of systematic errors, we found it 
worthwhile to determine the cumulants by a completely independent method, 
namely Monte Carlo simulations. For large enough system sizes the systematic
errors are negligible (of the order $\exp(-L/\xi)$, where $L$ is the linear
size of the system and $\xi$ the finite correlation length), and by 
increasing the simulation time also the statistical errors can be made as 
small as desired. In the following we report high-statistics measurements 
of the first ten energy cumulants in the ordered and disordered phase at 
$\beta_t$ for the models with $q=10$, 15, and 20. With increasing order
we observe the expected steep growth of the energy cumulants. The accuracy 
of the estimates, however, decreases with increasing order because the tail
ends of the energy distribution become more and more important.
Up to the third-order cumulant the precision of the numerical estimates
is still high (about 1\% or better), and we obtain very good agreement 
with the large-$q$ expansions down to $q=10$. At fourth order the agreement
between the two methods is still good within the statistical error bounds 
of about $5\%-15\%$, and also the results for the fifth and sixth order, 
the highest energy cumulants considered in Ref.~\cite{blm_JPI}, are 
compatible with each other. The further estimates up to the tenth order
should only be taken as a rough indication of the order of magnitude in 
high orders of the cumulant expansion. In addition we also present estimates
of the first three magnetization cumulants in the two phases at $\beta_t$, 
and compare the susceptibility in the ordered phase with recent 
low-temperature series expansions for $q=10$.

The remainder of the paper is organized as follows. In Sec.~2 we first
briefly recall the model and some exact results. We then describe the 
set-up of our simulations and discuss the estimators used for measuring
the various cumulants. The numerical data are presented in Sec.~3, and in
Sec.~4 we conclude with a brief summary of the main results and a few final
remarks.
%
           \section{Model and simulation}
%
In our simulations we employed the standard definition of 
the Potts model partition function \cite{potts,wu_potts}
\begin{equation}
  Z = e^{-\beta F} = \sum_{\{s_i\}} e^{-\beta E}; \, 
  E = -\sum_{\langle ij \rangle}
  \delta_{s_i s_j}; \, s_i = 1,\dots,q,
\label{eq:model}
\end{equation}
where $\beta = J/k_B T$ is the inverse temperature in natural units,
$F$ is the free energy,
$i$ denote the sites of a two-dimensional square lattice, 
$\langle ij \rangle$ are 
nearest-neighbor pairs, and $\delta_{s_i s_j}$ is the Kronecker delta symbol.

\begin{figure}[t]
\vskip 4.0truecm
\includegraphics{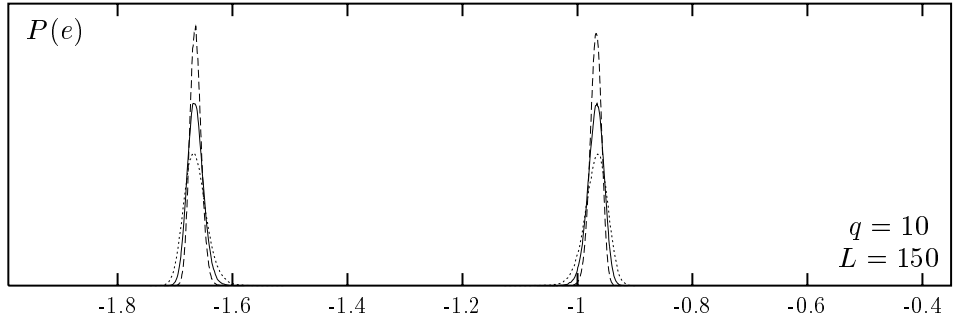}
\caption[a]{\label{fig:histo}%
The energy probability distribution $P(e)$ of the 10-state model at $\beta_t$ 
in the ordered and disordered phase for 
$L\times L$ (dotted lines), $2L\times L$ (solid lines), and
$2L \times 2L$ (dashed lines) lattices. The area under each peak 
is normalized to unity.}
\end{figure}

All simulations were performed in the canonical ensemble at 
$\beta = \beta_t = \ln(1 + \sqrt{q})$,
employing large square lattices of size $300 \times 300$, $120 \times 120$,
and $80 \times 80$ for $q=10$, 15, and 20, respectively, and
periodic boundary conditions. To stabilize the pure ordered or disordered
phase we took advantage of the extremely small tunneling probability for
large system sizes \cite{jk_lat93,jk_pla,jk_lat94,jk_epl}. Since the
tunneling proceeds via mixed phase configurations with
two interfaces, the probability is proportional to $\exp(-2\sigma_{od}L)$,
where $\sigma_{od}$ is the interface tension between the ordered and
disordered phase. For 2D Potts models $\sigma_{od}$ has been analytically
predicted \cite{bj92}, $2\sigma_{od} = 1/\xi_d(\beta_t)$, where
$\xi_d(\beta_t)$ is the exactly known \cite{xi} correlation length in the
disordered phase at the transition point ($ = 10.559519\dots$,
$4.180954\dots$, and $2.695502\dots$ for $q=10$, 15, and 20).
By rewriting the tunneling probability as $\exp(-L/\xi_d)$, it is easy to
estimate that for our choice of lattice sizes, $L \approx 28 \xi_d$,
the order of magnitude is about
$\exp(-28) \approx 10^{-12}$. It is therefore extremely probable that,
starting from a completely ordered or disordered configuration, the system
will stay in the ordered or disordered phase during a very long (but
finite) simulation time, thereby allowing statistically meaningful pure-phase
measurements of energy and magnetization cumulants. To be sure we have of
course monitored the
time evolution of our simulations and explicitly verified that no tunnelings
occurred. In Fig.~\ref{fig:histo} we
show the probability distributions $P(e)$ of the energy density $e=E/V$
in the ordered as well as in the disordered phase, demonstrating that
the two peaks are indeed very well separated.

The finite-size corrections in the pure phases are also expected to be of the
order $\exp(-L/\xi_p)$,
where the subscript $p$ stands for the ordered ($o$) and disordered ($d$)
phase, respectively. 
Since we have recently obtained strong numerical
evidence that $\xi_o(\beta_t) = \xi_d(\beta_t)$ \cite{jk_lat94,jk_epl},
this yields for the chosen lattice sizes in both phases again an order of
magnitude estimate of $\exp(-28) \approx 10^{-12}$.

To update the spins we employed in the ordered phase the heat-bath algorithm,
while in the disordered phase it is more efficient to use
the single-cluster algorithm \cite{jk_lat93,jk_pla}. The observed
integrated autocorrelation times of the energy, $\tau_{{\rm int},e}$,
and the statistics parameters are compiled in Table~\ref{tab:stat}. Notice
that for the heat-bath algorithm $\tau_{{\rm int},e}$ scales roughly with
$\xi_o^2$, as one would have expected on general grounds.
For the single-cluster algorithm we followed the usual convention and defined
$V/\langle |C| \rangle_{d, {\rm SC}}$ single-cluster steps as one Monte Carlo
update sweep (MCS), where $\langle |C| \rangle_{d, {\rm SC}}$ is the average
cluster size (cf.\ Table~\ref{tab:chi.dis} below).

The simulations with the heat-bath algorithm were performed on a CRAY vector
computer. To estimate statistical errors we divided the runs into
several bins and employed the standard jack-knife procedure.
The single-cluster code was implemented on a T3D parallel computer
by simulating 64 time histories in parallel. This enabled us to gather
an equivalent of about five workstation CPU years within a relatively short
time. Here the error bars are estimated from the fluctuations between the
64 copies by using again the jack-knife procedure.

The primary observables we report here are the energy cumulants 
$\kappa_p^{(n)}(\beta_t)$ at the transition point $\beta_t = \ln(1+\sqrt{q})$,
which are defined through the Taylor expansion of the scaled free energy
density $\beta f = \beta F/V$ around $\beta = \beta_t$,
\begin{equation}
-\beta f_p(\beta) = -\beta_t f(\beta_t) 
      + \sum_{n=1} (-1)^n \kappa_p^{(n)}(\beta_t) (\beta - \beta_t)^n/n!\,.
\label{eq:cumu_exp}
\end{equation}
While the free energy is continuous at a first-order
phase transition, $f(\beta_t) = f_o(\beta_t) = f_d(\beta_t)$, the cumulants
are discontinuous and hence different in the ordered and disordered phase.
The first three cumulants coincide with the (central) moments,
\begin{eqnarray}
\kappa_p^{(1)} &=& e_p = \langle E \rangle_p/V,\\
\kappa_p^{(2)} &=& c_p/\beta_t^2 = \mu_p^{(2)} 
=   ( \langle E^2 \rangle_p - \langle E \rangle_p^2)/V,\\
\kappa_p^{(3)} &=& \mu_p^{(3)} = 
  \langle (E - \langle E \rangle_p)^3 \rangle_p/V,
\end{eqnarray}
where $c_p$ is the usual specific heat in the pure phases.
The higher-order cumulants can be expressed as non-linear
combinations of the central moments 
$\mu_p^{(n)} = \langle (E - \langle E \rangle_p)^n \rangle_p/V$, e.g.,
\begin{eqnarray}
\kappa_p^{(4)} &=& \mu_p^{(4)} - 3V {\mu_p^{(2)}}^2,\\
\kappa_p^{(5)} &=& \mu_p^{(5)} - 10V \mu_p^{(2)} \mu_p^{(3)},\\
\kappa_p^{(6)} &=& \mu_p^{(6)} - 15V \mu_p^{(2)} \mu_p^{(4)}
                   -10V {\mu_p^{(3)}}^2 + 30V^2 {\mu_p^{(2)}}^3,
\label{eq:cumu_4_5}
\end{eqnarray}
and so on as listed up to the tenth order in the Appendix. While at $\beta_t$
both $e_o$ and $e_d$ are known exactly \cite{baxter}, the energy cumulants 
$\kappa_o^{(n)}$ and $\kappa_d^{(n)}$ with $n \ge 2$ can only be related to 
each other via duality \cite{wu_potts}. In particular for $c_o$ and $c_d$ 
as well as $\mu_o^{(3)}$ and $\mu_d^{(3)}$ the duality relations read
\begin{eqnarray}
c_o         &=& c_d - \beta^2_t (e_d - e_o )/\sqrt{q}, \label{eq:dual1}\\
\mu^{(3)}_o &=& -\mu^{(3)}_d +2(1-q)/q^{3/2} - 3(e_d - e_o)/q
                +6c_d/(\beta^2_t\sqrt{q}). \label{eq:dual2} 
\end{eqnarray}

For the magnetization cumulants we have used slightly different definitions
in the ordered and disordered phase. In the disordered phase the
magnetization vanishes and the magnetic susceptibility can be defined as
\begin{equation}
\chi_d = \frac{1}{V(q-1)^2}\langle \left(\sum_i (q \delta_{s_i,s} -1 ) 
         \right)^2 \rangle_d,
\label{eq:chi_d}
\end{equation}
which is certainly independent of the choice of the reference orientation 
$s = 1,\dots,q$. In simulations employing cluster-update algorithms the 
same quantity can also be estimated from
\begin{equation}
\chi_d = \frac{1}{q-1} \langle |C| \rangle_{d, {\rm SC}} 
       = \frac{1}{q-1} \frac{\langle |C|^2 \rangle_{d, {\rm SW}}}
                            {\langle |C| \rangle_{d, {\rm SW}}},
\label{eq:chi_d_clus}
\end{equation}
where $|C|$ denotes the size or weight of a cluster, i.e., the number of 
spins belonging to a cluster, and the subscripts ``SC'' and ``SW'' refer
to averages over the clusters encountered in the single-cluster and 
Swendsen-Wang formulation, respectively. Alternatively one could also measure
the (projected) spin-spin correlation function $g(i)$, using the spin or one
of the cluster representations, and derive $\chi_d$ as the ``integral'' of 
$g(i)$, $\chi_d = (q/(q-1)^2) \sum_i g(i)$ \cite{jk_lat93,jk_pla}. 

In the ordered phase we have measured the maximum definition of
the magnetization
\begin{equation}
M_{\rm max} = \frac{q N_{\rm max} - V}{q-1};\,\,\,\,\,
              N_{\rm max} = {\rm max}\{N_1,N_2,\dots,N_q\},
\label{eq:m_max}
\end{equation}
where $N_s$ counts the number of spins of orientation $s=1,\dots,q$ in a
given configuration.
A cluster estimator for the magnetization is 
\begin{equation}
M_{\rm clus} = |C_{\rm max}|,
\label{eq:m_clus}
\end{equation}
where $|C_{\rm max}|$ denotes the
size of the largest (spanning) stochastic cluster in each spin configuration. 
The expectation values $m = \langle M_{\rm max} \rangle_o/V$ and
$m' = \langle M_{\rm clus} \rangle_o/V$ coincide, $m = m'$, and can
be directly compared with the exact result for $q \ge 5$,
$m = \prod_{n=1}^{\infty} \left[ (1-x^n)/(1-x^{4n}) \right]$,
with $0<x<1$ defined by $q = x + 2 + x^{-1}$ \cite{baxter2}.
The magnetic susceptibility in the ordered phase is computed as
\begin{equation}
\chi_o = \left( \langle M_{\rm max}^2 \rangle_o - 
                \langle M_{\rm max} \rangle_o^2 \right)/V,
\label{eq:chi_o}
\end{equation}
and the third-order magnetization cumulant (or central moment) is given by
\begin{equation}
m_o^{(3)} = \langle \left( M_{\rm max} - 
            \langle M_{\rm max} \rangle_o \right)^3 \rangle_o/V.
\label{eq:m_o_3}
\end{equation}
Notice that while the expectation values of $M_{\rm max}$
and $M_{\rm clus}$ coincide, this is not the case for
the susceptibilities $\chi_o$ and $\chi'_o$, where the latter is 
defined by (\ref{eq:chi_o}) with $M_{\rm max}$ replaced
by $M_{\rm clus}$. The same remark applies of course also 
to the third-order cumulants $m_o^{(3)}$ and ${m'}_o^{(3)}$.

%
           \section{Results}
%
The Monte Carlo results for $e_p$, $c_p$, and $\mu_p^{(3)}$ at $\beta_t$ are
collected in Table~\ref{tab:dis_cumu} ($p=d$, disordered phase) and
Table~\ref{tab:ord_cumu} ($p=o$, ordered phase). For comparison we have also 
listed previous measurements \cite{jk_pla,jk_epl} on smaller lattices which 
clearly demonstrate that the above described ``dynamically stabilized''
pure phase simulations are feasible and that, as expected, the residual 
finite-size corrections are completely covered by the statistical errors.

The Table~\ref{tab:dis_cumu} shows the Monte Carlo results for the first
three energy cumulants at $\beta_t$ obtained from the simulations in the 
disordered phase. The first moments, $e_d$, are in excellent agreement with
the exact results, with relative statistical errors of the order 
$2 \times 10^{-5}$. This indicates quantitatively that systematic errors are
completely under control (including possible problems with pseudo-random 
numbers). The lines denoted by ``(large $q$)'' show for the readers 
convenience the duality transformed estimates from the large-$q$ expansions
for the cumulants in the ordered phase (cp.\ Table~\ref{tab:ord_cumu}). To 
be specific, for $\mu^{(3)}_d$ we have rewritten
(\ref{eq:dual2}) with the help of (\ref{eq:dual1}) as 
$\mu^{(3)}_d = -\mu^{(3)}_o +2(1-q)/q^{3/2} + 
3(e_d - e_o)/q + 6c_o/(\beta^2_t\sqrt{q})$ and inserted the large-$q$ 
estimates of Table~\ref{tab:ord_cumu}, which are taken from the most 
recent publication \cite{blm_JPI}. 

In Table~\ref{tab:ord_cumu} we show the first three energy cumulants at 
$\beta_t$ in the ordered phase obtained first directly from the simulations 
in the ordered phase (denoted by, e.g., ($o$, $L \times L$)), and second via
the duality relations (\ref{eq:dual1}) and (\ref{eq:dual2}) using the just 
described results in the disordered phase (denoted by, e.g., 
($d$, $L \times L$)) collected in Table~\ref{tab:dis_cumu}. Also in the 
ordered phase the exactly known first moments are confirmed with high 
precision, and the duality relations are very well satisfied by the two 
independent sets of Monte Carlo simulations, which further underlines the 
reliability of the data. As already mentioned above, the large-$q$ expansion
estimates are taken from Ref.~\cite{blm_JPI}. (Thanks to longer series 
expansions and constantly improved analysis techniques the numbers given in 
\cite{blm_epl,blm_lat93,blm_npb,blm_lat94,blm_JPI} scatter a little bit,
reflecting the current state of the art.) We see that in all cases the 
agreement between the Monte Carlo and large-$q$ estimates is extremely good. 
The only exception is perhaps $\mu^{(3)}_d$ for $q=10$, but here quite 
naturally the systematic error of the large-$q$ expansion is already 
relatively large. Recent analyses of low- and high-temperature series 
expansions for the $q=10$ Potts model specific heat at $\beta_t$,
on the other hand, yielded much larger values of $c_o = 31.8(2.8)$ and
$c_d = 33(3)$ \cite{beg94}.

\begin{figure}[htb]
\vskip 7.5truecm
\includegraphics{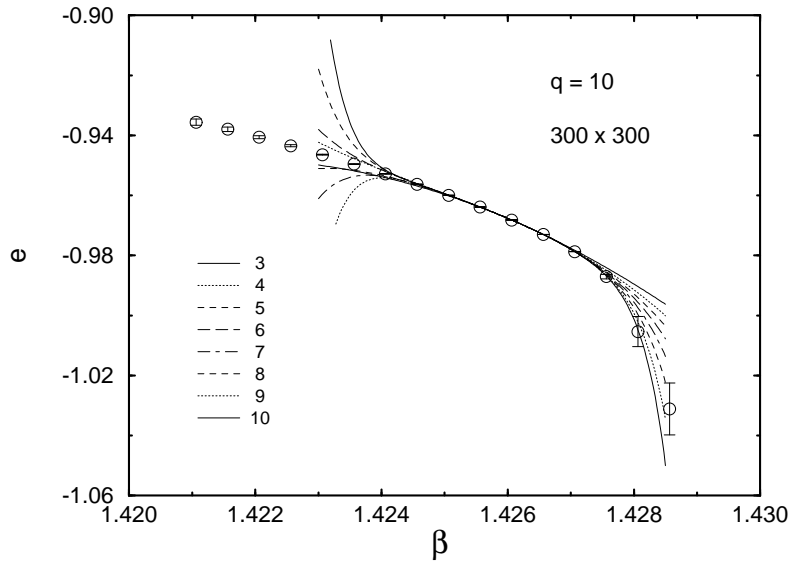}
\caption[]{\label{fig:e_cumu} Cumulant expansion 
of the energy density $e = -(d/d\beta) (-\beta f)$ for $q=10$ around
$\beta_t = \log(1+\sqrt{10}) \approx 1.426\,062\dots$ in the
disordered phase. The numbers $n=3,4,5,\dots$ in the legend indicate 
the highest 
order of the cumulants $\kappa_d^{(n)}$ involved in the expansion. The
open circles show for comparison the energy as obtained by standard
reweighting.}
\end{figure}

For the $2L \times L$ and $2L \times 2L$ lattices we have also computed 
higher-order energy moments and the resulting cumulants. Our results up to 
the eighth order are collected in Table~\ref{tab:higher_cumu}. With increasing
order these observables become very sensitive to the tail ends of the energy
distribution and the statistical accuracy deteriorates quite rapidly. Due to
cancellation effects this decrease in accuracy is much more pronounced for 
the cumulants than for the (central) moments. For $\kappa_p^{(4)}$ the 
statistical errors are about $5\%-15\%$, and here we still find quantitative
agreement of $\kappa_o^{(4)}$ with the resummed large-$q$ expansions, whose 
estimated systematic errors are of the same order. Also for $\kappa_o^{(5)}$ 
and $\kappa_o^{(6)}$ the agreement with the results read off from 
Fig.~13 of Ref.~\cite{blm_JPI} 
is quite satisfactorily, even though the statistical 
errors are obviously already quite large. For $\kappa_p^{(7)}$ and 
$\kappa_p^{(8)}$ some entries in Table~\ref{tab:higher_cumu} are no longer 
reliable and only the orders of magnitude should be trusted. Here we have 
certainly reached the limit of the present simulations, in particular for 
$q=10$,  and just as a very rough estimate we finally quote
$\kappa_o^{(9)} \approx 10^{23}$, $10^{19}$, and $10^{16}$, and 
$\kappa_o^{(10)}\approx 10^{27}$, $10^{22}$, and $10^{19}$, 
for $q=10$, 15, and 20, respectively. As an example we show in 
Fig.~\ref{fig:e_cumu} for $q=10$ the cumulant expansion around $\beta_t$ of
the energy density $e = -(d/d\beta) (-\beta f)$ in the disordered phase and
compare the results with extrapolations obtained by the standard reweighting
method.

The Table~\ref{tab:chi.dis} collects the expectation values for the
susceptibility $\chi_d$ at $\beta_t$, using the two different cluster 
estimators (\ref{eq:chi_d_clus}). For comparison we have also included 
the integral over the zero-momentum correlation function 
$g(i)$ \cite{jk_lat93,jk_pla}, which was also computed by employing
a Swendsen-Wang cluster estimator. One can show that this amounts only
to a different implementation of precisely the same operations needed
to compute directly the Swendsen-Wang cluster estimator in 
eq.~(\ref{eq:chi_d_clus}), and therefore the first and second lines for 
each lattice size in Table~\ref{tab:chi.dis} in fact turn out to be 
identical as they should.
 
The results in Table~\ref{tab:M.ord} for the magnetization in the ordered
phase clearly confirm that $m = m'$ with high precision. About 5-6 
significant digits of the numerical estimates agree within their $1\sigma$ 
error bounds with the exact results of Ref.~\cite{baxter2}, which provides
further evidence that all measured numbers can be interpreted as pure phase
expectation values. Furthermore we note that as expected the higher 
moments of $M_{\rm max}$ and $M_{\rm clus}$ do not agree. More
quantitatively we find consistently that 
$\chi_o > \chi'_o$ and $|m_o^{(3)}| > |{m'}_o^{(3)}|$.
The proper susceptibility $\chi_o$ for $q=10$ has also been considered in the 
low-temperature series analysis of Briggs {\em et al.\/} (BEG) \cite{beg94}. 
They obtained an estimate of $\chi(\beta_t^+)_{\rm BEG} = 2.44(9)$, 
leading to $\chi_o = (\frac{q}{q-1})^2 \chi(\beta_t^+)_{\rm BEG} = 3.01(12)$,
which is significantly smaller than
our value of $4.744(42)$ on the $300 \times 300$ lattice.
%
                         \section{Summary}
%
We have performed high-precision Monte Carlo simulations in the ordered and
disordered phase of 2D $q$-state Potts models with $q=10$, 15, and 20 
at their first-order transition point, working with large lattices of linear
size $L \approx 28 \xi_d$ ($= 300$, 120, 80). As an important 
self-consistency test the first three energy cumulants are found to satisfy
the duality relations with high precision, and both the energy and the 
magnetization are fully consistent with Baxter's exact values. As our main
result we obtain for the first three energy cumulants very good agreement with
recent resummed large-$q$ expansions of Bhattacharya 
{\em et al.\/} \cite{blm_JPI}, indicating that their technique can give
reliable results at least down to $q=10$. Also the fourth- to 
sixth-order cumulants are found in reasonably good agreement, albeit the
accuracy of both methods decreases with increasing order. We find, however,
significant differences to low- and high-temperature series analyses of 
the specific heat and magnetic susceptibility of the $q=10$ Potts model 
at $\beta_t$.
%
                         \section*{Acknowledgements}
%
WJ thanks the DFG for a Heisenberg fellowship and 
SK gratefully acknowledges a fellowship by the
Graduierten\-kolleg ``Physik and Chemie supra\-moleku\-larer Systeme''.
Work supported by computer grants hkf001 of HLRZ J\"ulich and
bvpf03 of Norddeutscher Vektorrechnerverbund (NVV) Berlin-Hannover-Kiel.
\clearpage\newpage
%
                         \section*{Appendix}
%
This appendix lists the relation between cumulants and central moments.
To simplify the notation we employ here the definitions
$\kappa_n = V \kappa_p^{(n)}$ and
$\mu_n = V \mu_p^{(n)} = \langle (E - \langle E \rangle_p)^n\rangle_p$.
With the help of computer algebra we obtained
\begin{eqnarray}
\kappa_4 &=& \mu_4 - 3 \mu_2^2, \nonumber\\
\kappa_5 &=& \mu_5 - 10 \mu_2 \mu_3, \nonumber\\
\kappa_6 &=& \mu_6 - 15 \mu_2 \mu_4 - 10 \mu_3^2 + 30 \mu_2^3, \nonumber\\
\kappa_7 &=& \mu_7 - 21 \mu_2 \mu_5 - 35 \mu_3 \mu_4 + 210 \mu_3 \mu_2^2, 
\nonumber \\
\kappa_8 &=& \mu_8 - 28 \mu_2 \mu_6 - 56 \mu_3 \mu_5 + 420 \mu_4 \mu_2^2 
                           - 35 \mu_4^2 
+ 560 \mu_3^2 \mu_2 - 630 \mu_2^4, \nonumber\\
\kappa_9 &=& \mu_9 - 36 \mu_2 \mu_7 - 84 \mu_3 \mu_6 - 126 \mu_4 \mu_5 
                           + 756 \mu_5 \mu_2^2 \nonumber\\
             & &           + \,2520 \mu_2 \mu_3 \mu_4 + 560 \mu_3^3
                           - 7560 \mu_3 \mu_2^3, \nonumber\\ 
\kappa_{10} &=& \mu_{10} - 45 \mu_2 \mu_8 - 120 \mu_3 \mu_7 - 210 \mu_4 \mu_6 
                             - 126 \mu_5^2 \nonumber\\
              & &            +\, 1260 \mu_6 \mu_2^2 + 5040 \mu_2 \mu_3 \mu_5 
                             + 3150 \mu_4^2 \mu_2 + 4200 \mu_4 \mu_3^2 \nonumber\\
              & &            -\, 18900  \mu_4 \mu_2^3 - 37800 \mu_3^2 \mu_2^2 
                             + 22680 \mu_2^5. \nonumber
\end{eqnarray}
%
%
\clearpage\newpage

\clearpage\newpage
%
%
\begin{table}[b]
\newlength{\digitwidth} \settowidth{\digitwidth}{\rm 0}
\catcode`?=\active \def?{\kern\digitwidth}
\caption[a]{Integrated autocorrelation time $\tau_{{\rm int},e}$ 
   of the energy and the number of Monte Carlo update sweeps (MCS) 
   in units of $\tau_{{\rm int},e}$.}
\label{tab:stat}
\begin{center}
\begin{tabular}{|l|l|l|l|}
\hline
                               &
 \multicolumn{1}{c|}{$q=10$}   &
 \multicolumn{1}{c|}{$q=15$}   &
 \multicolumn{1}{c|}{$q=20$}   \\
                               &
 \multicolumn{1}{c|}{$300\times300$}  &
 \multicolumn{1}{c|}{$120\times120$}  &
 \multicolumn{1}{c|}{$80\times80$}   \\
\hline
 \multicolumn{4}{|c|}{ordered phase (heat-bath algorithm)}\\
\hline
  $\tau_{{\rm int},e}$      & $\approx 170$ & $\approx 20$ & $\approx 9$ \\
  MCS/$\tau_{{\rm int},e}$  &      ?60\,000 &  ?\,640\,000 &  1\,280\,000 \\
\hline
 \multicolumn{4}{|c|}{disordered phase (single-cluster algorithm)}\\
\hline
 $\tau_{{\rm int},e}$     & $\approx 59$ & $\approx 18$ & $\approx  25$ \\
 MCS/$\tau_{{\rm int},e}$ &     600\,000 &  9\,000\,000 &   4\,200\,000 \\
\hline
\end{tabular}
\end{center}
\end{table}
%
%
\begin{table}[b]
\settowidth{\digitwidth}{\rm 0}
\catcode`?=\active \def?{\kern\digitwidth}
\caption[a]{Comparison of numerical and analytical results for energy
cumulants at $\beta_t$ in the disordered phase.}
\label{tab:dis_cumu}
\begin{center}
\begin{tabular}{|l|l|l|l|}
\hline
\multicolumn{1}{|c}{Observable} &
\multicolumn{1}{|c}{$q=10$} &
\multicolumn{1}{|c}{$q=15$} &
\multicolumn{1}{|c|}{$q=20$} \\
                               &
 \multicolumn{1}{c|}{$L=150$}  &
 \multicolumn{1}{c|}{$L=60$}  &
 \multicolumn{1}{c|}{$L=40$}   \\
\hline
 $e_d\quad$($d$, $\;\:L \times \;\;L$)      &$ -0.96812(15)  $&$ -0.75053(13)   $&$
-0.62648(20)  $\\
 $e_d\quad$($d$, $2L \times \;\;L$)         &$ -0.968190(81) $&$ -0.750510(65)  $&$
-0.626555(97) $\\
 $e_d\quad$($d$, $2L \times 2L$)            &$ -0.968186(18) $&$ -0.7504949(73) $&$
-0.626519(13) $\\
 $e_d\quad$(exact)                          &$ -0.968203...  $&$ -0.750492...   $&$
-0.626529...  $\\
\hline
 $c_d\quad$($d$, $\;\:L \times \;\;L$)      &$ 18.33(17)     $&$ 8.695(47)      $&$
6.144(43)     $\\
 $c_d\quad$($d$, $2L \times \;\;L$)         &$ 18.34(12)     $&$ 8.665(29)      $&$
6.140(27)     $\\
 $c_d\quad$($d$, $2L \times 2L$)            &$ 18.437(40)    $&$ 8.6507(57)     $&$
6.1327(38)    $\\
 $c_d\quad$(large $q$)                      &$ 18.43(2)      $&$ 8.657(3)       $&$
6.1326(4)     $\\
\hline
 $\mu_d^{(3)}\;$($d$, $\;\:L \times\;\; L$) &$ -2010(100)    $&$ -171.0(5.1)    $&$
-54.7(1.9)    $\\
 $\mu_d^{(3)}\;$($d$, $2L \times\;\; L$)    &$ -2031(73)     $&$ -176.1(3.8)    $&$
-53.9(1.5)    $\\
 $\mu_d^{(3)}\;$($d$, $2L \times  2L$)      &$ -2015(26)     $&$ -176.01(76)    $&$
-54.85(29)    $\\
 $\mu_d^{(3)}\;$(large $q$)                 &$ -1834(200)    $&$ -174(4)        $&$
-54.7(4)      $\\
\hline
\end{tabular}
\end{center}
\end{table}
%
%
\begin{table}[b]
\settowidth{\digitwidth}{\rm 0}
\catcode`?=\active \def?{\kern\digitwidth}
\caption[a]{Comparison of numerical and analytical results for energy
cumulants at $\beta_t$ in the ordered phase.}
\label{tab:ord_cumu}
\begin{center}
\begin{tabular}{|l|l|l|l|}
\hline
\multicolumn{1}{|c}{Observable} &
\multicolumn{1}{|c}{$q=10$} &
\multicolumn{1}{|c}{$q=15$} &
\multicolumn{1}{|c|}{$q=20$} \\
                               &
 \multicolumn{1}{c|}{$L=150$}  &
 \multicolumn{1}{c|}{$L=60$}  &
 \multicolumn{1}{c|}{$L=40$}   \\
\hline
 $e_o\quad$($o$, $\;\:L \times\;\; L$) &$ -1.664177(81) $&$ -1.765850(34) $&$ -1.820722(43) $\\
 $e_o\quad$($o$, $2L \times \;\;L$)    &$ -1.664262(57) $&$ -1.765875(26) $&$ -1.820689(14) $\\
 $e_o\quad$($o$, $2L \times 2L$)       &$ -1.664224(58) $&$ -1.765914(27) $&$ -1.820659(20) $\\
 $e_o\quad$(exact)                     &$ -1.664253...  $&$ -1.765906...  $&$ -1.820684...  $\\
\hline
 $c_o\quad$($o$, $\;\:L \times\;\; L$) &$ 17.95(13)  $&$ 8.016(21)  $&$ 5.351(15)  $\\
 $c_o\quad$($d$, $\;\:L \times\;\; L$) &$ 17.88(17)  $&$ 8.037(47)  $&$ 5.373(43)  $\\
 $c_o\quad$($o$, $2L \times \;\;L$)    &$ 17.81(10)  $&$ 8.004(19)  $&$ 5.3612(55) $\\
 $c_o\quad$($d$, $2L \times \;\;L$)    &$ 17.89(12)  $&$ 8.007(29)  $&$ 5.369(27)  $\\
 $c_o\quad$($o$, $2L \times 2L$)       &$ 18.00(10)  $&$ 7.990(19)  $&$ 5.3608(88) $\\
 $c_o\quad$($d$, $2L \times 2L$)       &$ 17.989(40) $&$ 7.9931(57) $&$ 5.3613(38) $\\
 $c_o\quad$(large $q$)                 &$ 17.98(2)   $&$ 7.999(3)   $&$ 5.3612(4)  $\\
\hline
 $\mu_o^{(3)}\;$($o$, $\;\:L \times\;\; L$) &$ 1979(87)  $&$ 180.5(3.1)   $&$ 57.0(1.3)  $\\
 $\mu_o^{(3)}\;$($d$, $\;\:L \times\;\; L$) &$ 2026(100) $&$ 175.7(5.1)   $&$ 56.9(1.9)  $\\
 $\mu_o^{(3)}\;$($o$, $2L \times \;\;L$)    &$ 1836(71)  $&$ 189.7(5.1)   $&$ 56.24(40)  $\\
 $\mu_o^{(3)}\;$($d$, $2L \times \;\;L$)    &$ 2047(73)  $&$ 180.8(3.8)   $&$ 56.1(1.5)  $\\
 $\mu_o^{(3)}\;$($o$, $2L \times 2L$)       &$ 2030(110) $&$ 177.2(3.2)   $&$ 56.83(71)  $\\
 $\mu_o^{(3)}\;$($d$, $2L \times 2L$)       &$ 2031(26)  $&$ 180.67(76)   $&$ 57.09(29)  $\\
 $\mu_o^{(3)}\;$(large $q$)                 &$ 1900(200) $&$ 179(4)       $&$ 56.9(4)    $\\
\hline
\end{tabular}
\end{center}
\end{table}
%
%
\begin{table}[b]
\settowidth{\digitwidth}{\rm 0}
\catcode`?=\active \def?{\kern\digitwidth}
\caption[a]{Numerical estimates of higher-order energy
cumulants in the disordered and ordered phase at $\beta_t$.}
\label{tab:higher_cumu}
\begin{center}
\begin{tabular}{|l|l|l|l|}
\hline
\multicolumn{1}{|c}{Observable} &
\multicolumn{1}{|c}{$q=10$} &
\multicolumn{1}{|c}{$q=15$} &
\multicolumn{1}{|c|}{$q=20$} \\
                               &
 \multicolumn{1}{c|}{$L=150$}  &
 \multicolumn{1}{c|}{$L=60$}  &
 \multicolumn{1}{c|}{$L=40$}   \\
\hline
 $\kappa_d^{(4)}\;$($d$, $2L \times \;\;L$) &$ 1.61(14)? \times 10^6    $&$ 3.07(24)? \times 10^4     $&$ 4.87(38)? \times 10^3     $\\
 $\kappa_d^{(4)}\;$($d$, $2L \times 2L$)    &$ 1.583(64) \times 10^6    $&$ 3.193(48) \times 10^4     $&$ 4.905(89) \times 10^3     $\\
 $\kappa_o^{(4)}\;$($o$, $2L \times \;\;L$) &$ 1.10(13)? \times 10^6    $&$ 3.95(47)? \times 10^4     $&$ 4.67(14)? \times 10^3     $\\
 $\kappa_o^{(4)}\;$($o$, $2L \times 2L$)    &$ 1.55(22)? \times 10^6    $&$ 2.93(20)? \times 10^4     $&$ 4.79(22)? \times 10^3     $\\
$\kappa_o^{(4)}\;$(large $q$)                  &$ 1.3(2)??? \times 10^6    $&$ 3.1(2)??? \times 10^4     $&$ 5.0(1)??? \times 10^3 $\\
\hline
 $\kappa_d^{(5)}\;$($d$, $2L \times \;\;L$) &$ -2.39(39) \times 10^9    $&$ -1.10(23)?  \times 10^7   $&$ -8.1(1.1)\,  \times 10^5    $\\
 $\kappa_d^{(5)}\;$($d$, $2L \times 2L$)    &$ -2.40(26) \times 10^9    $&$ -1.170(59)  \times 10^7   $&$ -8.42(32)  \times 10^5    $\\
 $\kappa_o^{(5)}\;$($o$, $2L \times \;\;L$) &$ ?\,\,1.03(25) \times 10^9   $&$ ?\,\,1.73(49)? \times 10^7   $&$ ?\,\,7.16(65) \times 10^5    $\\
 $\kappa_o^{(5)}\;$($o$, $2L \times 2L$)    &$ ?\,\,1.98(51) \times 10^9   $&$ ?\,\,0.92(15)? \times 10^7   $&$ ?\,\,7.37(70) \times 10^5    $\\
$\kappa_o^{(5)}\;$(large $q$)                  &$ ?????\approx 10^9    $&$ ?????\approx 10^7     $&$ ?????\approx 10^6 $\\
\hline
 $\kappa_d^{(6)}\;$($d$, $2L \times \;\;L$) &$ 5.1(1.2)\, \times 10^{12}  $&$ ?6.6(2.7) \times 10^9    $&$ 1.77(34) \times 10^8     $\\
 $\kappa_d^{(6)}\;$($d$, $2L \times 2L$)    &$ 5.8(1.3)\, \times 10^{12}  $&$ ?7.2(1.1) \times 10^9    $&$ 2.33(16) \times 10^8     $\\
 $\kappa_o^{(6)}\;$($o$, $2L \times \;\;L$) &$ 1.13(48)  \times 10^{12}  $&$  11.9(5.4)  \times 10^9 $&$ 1.63(36) \times 10^8     $\\
 $\kappa_o^{(6)}\;$($o$, $2L \times 2L$)    &$ 2.8(1.4)\, \times 10^{12}  $&$ ?4.5(1.3)  \times 10^9   $&$ 1.38(26) \times 10^8     $\\
$\kappa_o^{(6)}\;$(large $q$)                  &$ ?????\approx 10^{12}    $&$ ???\approx 3 \times 10^9     $&$ ???\approx 2 \times 10^8 $\\
\hline
 $\kappa_d^{(7)}\;$($d$, $2L \times \;\;L$) &$ -1.20(39) \times 10^{16} $&$ -5.5(3.5)\, \times 10^{12} $&$ -4.0(1.1) \,\times 10^{10} $\\
 $\kappa_d^{(7)}\;$($d$, $2L \times 2L$)    &$ -2.01(66) \times 10^{16} $&$ -6.8(2.1)\, \times 10^{12} $&$ -8.47(96) \times 10^{10} $\\
 $\kappa_o^{(7)}\;$($o$, $2L \times \;\;L$) &$ ?\,\,1.15(92) \times 10^{15} $&$ ?\,\,1.03(60) \times 10^{13}  $&$ ?\,\,5.0(2.2) \,\times 10^{10}  $\\
 $\kappa_o^{(7)}\;$($o$, $2L \times 2L$)    &$ ?\,\,3.9(4.2)\, \times 10^{15} $&$ ?\,\,3.0(1.2)\, \times 10^{12}  $&$ ?\,\,2.6(1.0)\,\times 10^{10}   $\\
\hline
 $\kappa_d^{(8)}\;$($d$, $2L \times \;\;L$) &$ 2.9(1.3) \times 10^{19}  $&$ 5.5(4.4)\, \times 10^{15}  $&$ 7.9(3.7)\, \times 10^{12}  $\\
 $\kappa_d^{(8)}\;$($d$, $2L \times 2L$)    &$ 8.6(3.5) \times 10^{19}  $&$ 8.8(4.4)\, \times 10^{15}  $&$ 3.57(56) \times 10^{13}  $\\
 $\kappa_o^{(8)}\;$($o$, $2L \times \;\;L$) &$ ?\,\, \approx 1  \times 10^{17}           $&$ 1.01(69) \times 10^{16}  $&$ 2.0(1.3)\, \times 10^{13}  $\\
 $\kappa_o^{(8)}\;$($o$, $2L \times 2L$)    &$ \approx -2 \times 10^{15}        $&$ 2.1(1.1)\, \times 10^{15}  $&$ 3.1(4.2)\, \times 10^{12}  $\\
\hline
\end{tabular}
\end{center}
\end{table}
%
%
%
\begin{table}[h]
\caption[a]
      {\label{tab:chi.dis}
      The magnetic susceptibility $\chi_d$ at $\beta_t$ in the disordered phase
      using different estimators.}
\begin{center}
\begin{tabular}{|c|l|l|l|l|}
\hline
 \multicolumn{1}{|c}{Lattice}       &
 \multicolumn{1}{|c}{Observable}       &
 \multicolumn{1}{|c}{$q=10$} &
 \multicolumn{1}{|c}{$q=15$} &
 \multicolumn{1}{|c|}{$q=20$} \\
                &               &
 \multicolumn{1}{c|}{$L=150$}  &
 \multicolumn{1}{c|}{$L=60$}  &
 \multicolumn{1}{c|}{$L=40$}   \\
\hline
  &$\frac{q}{(q-1)^2}\sum^{L}_{i=1}g(i)                                              $&$ 4.224(16)  $&$ 0.7306(14)   $&$ 0.3092(58)  $\\
$L\times L$&$\frac{1}{q-1} \langle|C|^2\rangle_{d, {\rm SW}}/\langle|C|\rangle_{d, {\rm SW}}   $&$ 4.224(16)  $&$ 0.7306(14)   $&$ 0.3092(58)  $\\
  &$\frac{1}{q-1} \langle |C|   \rangle_{d, {\rm SC}}                                     $&$ 4.224(16)  $&$ 0.7310(14)   $&$ 0.3090(58)  $\\
\hline
  &$\frac{q}{(q-1)^2}\sum^{2L}_{i=1}g(i)                                             $&$ 4.2306(89) $&$ 0.73093(68)  $&$ 0.30954(32) $\\
$2L\times L$&$\frac{1}{q-1} \langle|C|^2\rangle_{d, {\rm SW}}/\langle|C|\rangle_{d, {\rm SW}}   $&$ 4.2306(89) $&$ 0.73093(68)  $&$ 0.30954(32) $\\
  &$\frac{1}{q-1} \langle |C|   \rangle_{d, {\rm SC}}                                     $&$ 4.2327(89) $&$ 0.73094(65)  $&$ 0.30952(31) $\\
\hline
  &$\frac{q}{(q-1)^2}\sum^{2L}_{i=1}g(i)                                             $&$ 4.2326(18) $&$ 0.730386(79) $&$ 0.309356(39) $\\
$2L\times 2L$&$\frac{1}{q-1} \langle|C|^2\rangle_{d, {\rm SW}}/\langle|C|\rangle_{d, {\rm SW}} $&$ 4.2326(18) $&$ 0.730386(79) $&$ 0.309356(39) $\\
\hline
\end{tabular}
\end{center}
\end{table}
%
%
%
\begin{table}[hb]
\caption[a]{\label{tab:M.ord}
The magnetization $m$ and $m'$ at $\beta_t$ in the ordered phase, 
using the two estimators $M_{\rm max}$ and $M_{\rm clus}$, and the
corresponding susceptibilities and third moments.}
\begin{center}
\begin{tabular}{|l|l|l|l|}
\hline
 \multicolumn{1}{|c}{Observable}       &
 \multicolumn{1}{|c}{$q=10$} &
 \multicolumn{1}{|c}{$q=15$} &
 \multicolumn{1}{|c|}{$q=20$} \\
                               &
 \multicolumn{1}{c|}{$L=150$}  &
 \multicolumn{1}{c|}{$L=60$}  &
 \multicolumn{1}{c|}{$L=40$}   \\
\hline
 $m\:\quad$($\;\:L \times \;\;L$) &$ 0.857047(71) $&$ 0.916631(21) $&$ 0.941199(21)  $\\
 $m' \quad$($\;\:L \times \;\;L$) &$ 0.857047(71) $&$ 0.916634(21) $&$ 0.941197(21)  $\\
 $m\:\quad$($2L \times \;\;L$)    &$ 0.857113(49) $&$ 0.916648(16) $&$ 0.9411782(66) $\\
 $m' \quad$($2L \times \;\;L$)    &$ 0.857113(49) $&$ 0.916648(16) $&$ 0.9411791(66) $\\
 $m\:\quad$($2L \times    2L$)    &$ 0.857081(49) $&$ 0.916672(15) $&$ 0.9411694(97) $\\
 $m' \quad$($2L \times    2L$)    &$ 0.857077(49) $&$ 0.916672(15) $&$ 0.9411694(97) $\\
 $m\:\quad$(exact)                    &$ 0.857106...  $&$ 0.916663...  $&$ 0.9411759...  $\\
\hline
 $\chi_o  \quad$($\;\:L \times \;\;L$)   &$ 4.750(60) $&$ 0.8090(36) $&$ 0.3348(17)  $\\
 $\chi'_o \quad$($\;\:L \times \;\;L$)   &$ 4.704(60) $&$ 0.7989(36) $&$ 0.3305(17)  $\\
 $\chi_o  \quad$($2L \times \;\;L$)      &$ 4.663(43) $&$ 0.8095(38) $&$ 0.33509(55)  $\\
 $\chi'_o \quad$($2L \times \;\;L$)      &$ 4.623(43) $&$ 0.7997(38) $&$ 0.33076(55)  $\\
 $\chi_o  \quad$($2L \times    2L$)      &$ 4.744(42) $&$ 0.8052(28) $&$ 0.33551(81)  $\\
 $\chi'_o \quad$($2L \times    2L$)      &$ 4.700(43) $&$ 0.7953(28) $&$ 0.33118(80)  $\\
\hline
 ${m_o^{(3)}}\:$($\;\:L \times \;\;L$) &$ -1521(85) $&$ -45.9(1.2) $&$ -8.55(32) $\\
 ${m'}_o^{(3)} $($\;\:L \times \;\;L$) &$ -1505(84) $&$ -45.4(1.2) $&$ -8.44(32) $\\
 ${m_o^{(3)}}\:$($2L \times \;\;L$)    &$ -1372(62) $&$ -49.4(2.2) $&$ -8.321(88)  $\\
 ${m'}_o^{(3)} $($2L \times \;\;L$)    &$ -1362(62) $&$ -48.9(2.2) $&$ -8.216(88)  $\\
 ${m_o^{(3)}}\:$($2L \times    2L$)    &$ -1532(74) $&$ -45.3(1.0) $&$ -8.44(13)  $\\
 ${m'}_o^{(3)} $($2L \times    2L$)    &$ -1517(75) $&$ -44.8(1.0) $&$ -8.33(13)  $\\
\hline
\end{tabular}
\end{center}
\end{table}
\clearpage\newpage
%
%
%
%
\end{document}